\documentclass[a4paper,12pt]{article}
\usepackage{graphicx}
\usepackage{graphics}
\usepackage{amsmath}
\usepackage{geometry}
\usepackage{setspace}
\usepackage[square,numbers,sort&compress]{natbib}
\title{Restricted three body problems at the nanoscale}
\author{Yue Chan\footnote{Corresponding author, email-address: yc321@uow.edu.au}, Ngamta Thamwattana and James M Hill\\
\footnotesize{Nanomechanics Group, School of Mathematics and Applied Statistics,}\\
\footnotesize{University of Wollongong, Wollongong, NSW 2522,
Australia}} 
\begin{document}

\maketitle

\abstract{In this paper, we investigate some of the classical
restricted three body problems at the nanoscale, such as the
circular planar restricted problem for three C$_{60}$ fullerenes,
and a carbon atom and two C$_{60}$ fullerenes. We model the van der
Waals forces between the fullerenes by the Lennard-Jones potential.
In particular, the pairwise potential energies between the carbon
atoms on the fullerenes are approximated by the continuous approach,
so that the total molecular energy between two fullerenes can be
determined analytically. Since we assume that such interactions
between the molecules occur at sufficiently large distance, the
classical three body problems analysis is legitimate to determine
the collective angular velocity of the two and three C$_{60}$
fullerenes at the nanoscale. We find that the maximum angular
frequency of the two and three fullerenes systems reach the
terahertz range and we determine the stationary points and the
points which have maximum velocity for the carbon atom for the
carbon atom and the two fullerenes system.
\bigskip


}

\section{Introduction}
In 1966, Jones \cite{Jones} suggested in the New Scientist under the
name of Daedalus that the hollow all-carbon cage molecules were
possible. However, it was not until two decades later that Kroto
\cite{Smalley} experimentally discovered C$_{60}$ and C$_{70}$
fullerenes by analyzing the resulting mass spectrometry. The
dominant peak represents C$_{60}$ fullerene followed by a second
peak representing C$_{70}$ fullerene. In particular, C$_{60}$
fullerene has I$_{h}$ symmetry, which has a spheroidal shape.

Various nanoscale gigahertz oscillators have been proposed in the
recent literature including the oscillation of a C$_{60}$ fullerene
or a closed carbon nanotube inside an open ended carbon nanotube.
Such oscillators have also been confirmed by experiments on
multi-walled carbon nanotubes. Cumings and Zettl \cite{Cumings}
remove the cap from one end of the outer shell and attach a moveable
nanomanipulator to the core of a multi-walled nanotube through a
high-resolution transmission electron microscope. They observe that
the extruded core, after release, quickly and fully retracts inside
the outer shell due to the restoring force resulting from the van
der Waals interactions acting on the extruded core. This
experimental result lead to the molecular dynamics studies of Zheng
and Jiang \cite{Zheng} who show that the oscillating of the inner
shell between the open ends of the outer shell of a multi-walled
carbon nanotube generates a gigahertz frequency. Molecular dynamics
simulations on gigahertz oscillators have also been undertaken by
Legoas \emph{et al.} \cite{Legoas} and Rivera \emph{et al.}
\cite{Rivera1,Rivera2}. In terms of mathematical modelling, Baowan
and Hill \cite{Baowan} investigate the force distribution for a
double-walled carbon nanotube oscillator by adopting the continuous
approach for the Lennard-Jones potential together with Newton's
second law, assuming a frictionless environment, to investigate the
associated mechanics. They obtain an analytical expression for the
interaction force and their model also predicts the gigahertz
oscillatory behavior for the double-walled carbon nanotube
oscillators.

Based on the molecular dynamics simulations of various nano
gigahertz oscillators, Cox \emph{et al.} \cite{Cox,Cox2} develop an
accurate mathematical model employing fundamental mechanical
principles and classical applied mathematical techniques to
determine an acceptance condition and the suction energies of a
C$_{60}$ fullerene entering a nanotube. They then determine the
minimum radius of a carbon nanotube for which the C$_{60}$ fullerene
will be accepted from rest and the maximum total kinetic energy once
the C$_{60}$ molecule is sucked inside the nanotube by the van der
Waals forces. In addition, Cox \emph{et al.} \cite{Cox,Cox2} show
that the gigahertz oscillatory behavior arises from the two
peak-like forces operating at the nanotube's open ends. The
analytical model of Cox \emph{et al.} \cite{Cox,Cox2} is also
extended to examine some more complicated structures of the
gigahertz oscillators, including the nanotube bundle oscillators,
for which a single nanotube or a C$_{60}$ fullerene oscillates
inside the bundle \cite{Cox3,Cox4}. For other types of nanoscale
oscillators, Hilder and Hill \cite{Hilder2,Hilder3} find that the
gigahertz frequencies can also be obtained from a sector of a
nanotube orbiting inside a carbon nanotorus and an atom and a
C$_{60}$ fullerene orbiting inside a nanotorus.

Chan \emph{et al.} \cite{Chan} have investigated various two body
problem, namely fullerene-fullerene, fullerene-carbon nanotube at
the nanoscale. In this paper, we investigate some of the classical
restricted three body problems at the nanoscale. In particular, we
consider three C$_{60}$ fullerenes interacting with each other and
initially located at the vertices of an equilateral triangle. The
van der Waals interaction energy is modelled using the 6-12
Lennard-Jones potential and the continuous approach for which we
assume a uniform distribution of carbon atoms on the surfaces of the
carbon nanotube and the fullerenes. We determine the motion of the
interacting molecules as they move relatively with respect to each
other under the influence of their mutual centripetal force. For the
case under consideration, the circular orbiting frequencies reach
the gigahertz range. We comment that this paper ignores any thermal
fluctuations arising from the environment.

In section 2, we develop the mathematical model for the restricted
three body problems at the nanoscale. In section 3, we discuss the
restricted problem for the three fullerenes system. In section 4, we
investigate the capture of a carbon atom by two fullerenes and we
present some conclusions in the final section.

\section{Theory}
\begin{figure}
  \centering
  \includegraphics[width=10cm,height=10cm]{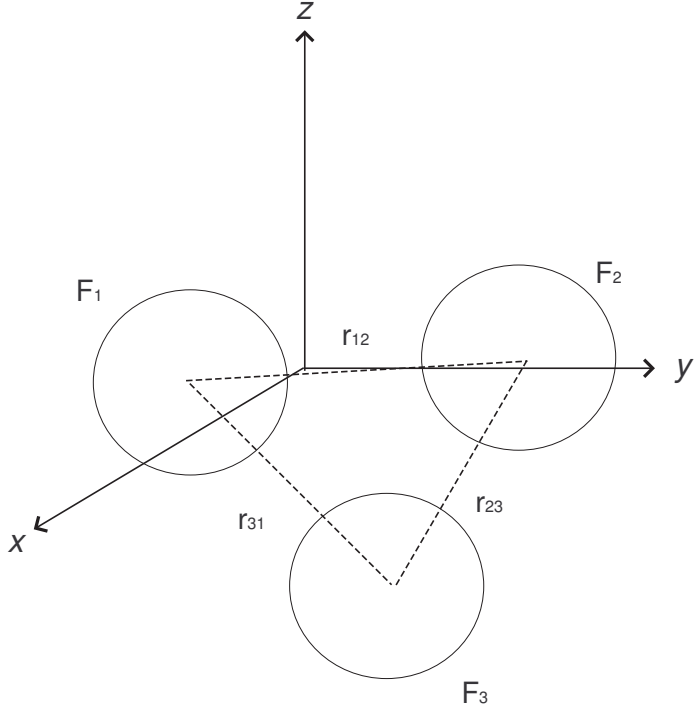}\\
  \caption{Three body problems at the nanoscale. Fullerene centers initially lying at the vertices of an equilateral triange}\label{fig:Fig1}
\end{figure}

In this section, we derive the theoretical basis for the study of
the three body problems, namely that of three C$_{60}$ fullerenes at
the nanoscale (see Fig.~\ref{fig:Fig1} for details). Since we only
consider the residual molecular interactions between molecules, it
is valid to examine the classical three body problems to investigate
the same problem at the nanoscale
\cite{Poole,Christian,Pollard,Moulton,Freundlich,Sterne,Mccuskey}.
We model the residual molecular interactions by van der Waals
forces, and in particular the Lennard-Jones potential
\cite{Lennard}, which is given by

\begin{equation}
V(\rho)=4\varepsilon\left[\left(\frac{\sigma}{\rho}
\right)^{12}-\left( \frac{\sigma}{\rho}\right)^{6}
\right]=-\frac{A}{\rho^{6}}+\frac{B}{\rho^{12}},\label{2.1}
\end{equation}
where $\rho$, $\varepsilon$ and $\sigma$ denote the distance between
two arbitrary carbon atoms, the potential well depth of two carbon
atoms and the parameter that is determined by the equilibrium
distance respectively. In addition, $A$ and $B$ denote the
attractive and the repulsive constants respectively. We assume that
the carbon atoms are evenly distributed over the surfaces of the
molecules such that the pairwise molecular energy between two
fullerenes, i.e $\sum_{k}\sum_{j}V(\rho_{kj})$ can be approximated
by the continuous approach, which has been previously used by Cox
\emph{et al.} \cite{Cox,Cox2}. That is,

\begin{equation}
V(r)=n_{f}^{2}\int_{S_{1}}\int_{S_{2}}\left(-\frac{A}{\rho^{6}}+\frac{B}{\rho^{12}}\right)
dS_{2}dS_{1},\label{2.2}
\end{equation}
where $n_{f}$, $\rho$, $dS_{1}$ and $dS_{2}$ denote the atomic
number density, the distance between the centers of two fullerenes,
the surface area element of the two fullerenes respectively.
Following the same calculation as that derived by Cox \emph{et
al}\cite{Cox,Cox2}, we obtain

\begin{equation}
V(r)=-Q_{6}(r)+Q_{12}(r),\label{2.3}
\end{equation}
where $Q_{n}$ is defined by

\begin{displaymath}
Q_{n}(r)=\frac{4\pi^{2}a^{2}C_{n}n_{f}^{2}}{r(n-2)(n-3)}\left\{\frac{1}{(2a+r)^{n-3}}
-\frac{1}{(2a-r)^{n-3}}-\frac{2}{r^{n-3}} \right\},
\end{displaymath}
where $a$, $C_{6}$ and $C_{12}$ denote the radius of the C$_{60}$
fullerene, A and B respectively. Given that, the mutual force,
$F(r)$ between two fullerenes can be determined by

\begin{equation}
F(r)=-\frac{dV(r)}{dr},\label{2.4}
\end{equation}
where $d /dr$ denotes the derivative in the $r$-direction.

According to Newton's second law, we can determine the forces acting
on a fullerene by the other two fullerenes as

\begin{equation}
m_{k}\ddot{\vec{r_{k}}}=\sum_{j\neq
k=1}^{3}F(r_{kj})\frac{\vec{r_{j}}-\vec{r_{k}}}{r_{kj}},\label{2.5}
\end{equation}
where $m_{k}$, $r_{k}$, $r_{kj}$ denote the total mass of the
$k$-fullerene, the position vector of the $k$-fullerene and the
relative displacement of any two fullerenes $k$ and $j$. We note
that Eqs.~(\ref{2.5}) can not be generally solved analytically. If
we make $\vec{r}=\vec{r_{2}}-\vec{r_{1}}$ and we assume that
$\vec{R}$ denotes the center of mass of $m_{1}$ and $m_{2}$
allocated at the origin of the coordinate system. We further define
$\vec{\rho}=\vec{r_{3}}-\vec{R}=M\mu^{-1}\vec{r_{3}}$, where
$\mu=m_{1}+m_{2}$. Under these assumptions, we can write these
relative displacements as

\begin{equation}
\vec{r_{2}}-\vec{r_{1}}=\vec{r}, \quad\quad
\vec{r_{3}}-\vec{r_{1}}=\vec{\rho}+m_{2}\mu^{-1}\vec{r}, \quad\quad
\vec{r_{3}}-\vec{r_{2}}=\vec{\rho}-m_{1}\mu^{-1}\vec{r}.\label{2.6}
\end{equation}
The equations of motion in terms of the Jacobi coordinates,
$\vec{r}$ and $\vec{\rho}$ are then given by

\begin{eqnarray}
&&
\ddot{\vec{r}}=-\frac{\mu}{m_{1}m_{2}}\frac{F(r_{12})}{r_{12}}\vec{r}+\frac{1}{m_{2}}\frac{F(r_{23})}{r_{23}}(\vec{\rho}-m_{1}\mu^{-1}\vec{r})-\frac{1}{m_{1}}\frac{F(r_{13})}{r_{13}}(\vec{\rho}+m_{1}\mu^{-1}\vec{r}),
\nonumber\\
&&
\ddot{\vec{\rho}}=-\frac{M}{m_{3}}\mu^{-1}\left[\frac{F(r_{13})}{r_{13}}(\vec{\rho}+m_{2}\mu^{-1}\vec{r})+\frac{F(r_{23})}{r_{23}}(\vec{\rho}-m_{1}\mu^{-1}\vec{r})\right],\label{2.7}
\end{eqnarray}

\section{Three identical C$_{60}$ fullerenes}

We now make the following three assumptions, referred to as the
circular planar restricted problem, to determine the motion
analytically:

\begin{enumerate}
  \item The three fullerenes are moving uniformly in circular
  orbits
  \item They are orbiting collectively in the same plane
  \item Their angular velocities are the same
\end{enumerate}

Upon assuming the three fullerenes are moving in plane, we can
denote $(x_{i},y_{i},0)$ by the coordinates of the $i$-fullerene.
Then, the second derivative of $x_{i}$ and $y_{i}$ can be written as

\begin{equation}
\ddot{x_{k}}=m_{k}^{-1}\sum_{j\neq
k}\frac{F(r_{jk})}{r_{jk}}(x_{j}-x_{k}), \quad\quad
\ddot{y_{k}}=m_{k}^{-1}\sum_{j\neq
k}\frac{F(r_{jk})}{r_{jk}}(y_{j}-y_{k}). \label{2.8}
\end{equation}
We now assume that the three fullerenes have the same (collective)
angular velocity $\omega$, and they are moving uniformly in a
circle, such that we can write down the rotating coordinate system

\begin{equation}
x_{k}=\xi_{k}\cos\omega t-\eta_{k}\sin\omega t.\label{2.9}
\end{equation}
We note that $y_{k}$ can be defined in a similar way as $x_{k}$. We
then substitute Eq.~(\ref{2.9}) into Eqs.~(\ref{2.8})$_{1}$ to
obtain

\begin{eqnarray}
&&
\ddot{\xi_{k}}-2\omega\dot{\eta_{k}}-\omega^{2}\xi_{k}=m_{k}^{-1}\sum_{j\neq
k}\frac{F(r_{jk})}{r_{jk}}(\xi_{j}-\xi_{k}),\nonumber\\
&&\ddot{\eta_{k}}+2\omega\dot{\xi_{k}}-\omega^{2}\eta_{k}=m_{k}^{-1}\sum_{j\neq
k}\frac{F(r_{jk})}{r_{jk}}(\eta_{j}-\eta_{k}).\label{2.10}
\end{eqnarray}
We can then recast Eq.~(\ref{2.10}) into a single equation by using
$z_{k}=\xi_{k}+ i \eta_{k}$, where $i$ denotes the usual imaginary
unit $\sqrt{-1}$, so that,

\begin{equation}
\ddot{z_{k}}+i2\omega\dot{z_{k}}-\omega^{2}z_{k}=m_{k}^{-1}\sum_{j\neq
k}\frac{F(r_{jk})}{r_{jk}}(z_{j}-z_{k}).\label{2.11}
\end{equation}
Since the fullerenes appear to be at rest in the rotating frame, we
can simplify Eq.~(\ref{2.11}) to become

\begin{equation}
-z_{k}=\lambda_{k}\sum_{j\neq
i}\frac{F(r_{jk})}{r_{jk}}(z_{j}-z_{k}), \label{2.12}
\end{equation}
where $\lambda_{k}=1/(m_{k}\omega^{2})$. If we let
$\beta_{j}=\lambda_{k}(F(r_{m\ell})/r_{m\ell})$, where $m,\ell \neq
j$, then Eq.~(\ref{2.12}) gives the following a system of three
linear equations

\begin{eqnarray}
&& \left\{
\begin{array}{rcl}(1-\beta_{3}-\beta_{2})z_{1}+\beta_{3}z_{2}+\beta_{2}z_{3}&=&0,\\
\\
\beta_{2}z_{1}+\beta_{1}z_{2}+(1-\beta_{2}-\beta_{1})z_{3}&=&0,\\
\\
m_{1}z_{1}+m_{2}z_{2}+m_{3}z_{3}&=&0,\\\end{array}\right.
\label{2.13}
\end{eqnarray}

For the three fullerenes moving in the same plane, geometrically,
two possibilities arise, namely; the three fullerenes are not on the
same straight line or they lie on the same straight line. For the
first scenario, the coefficients of $z_{j}$ of
Eqs.~(\ref{2.13})$_{1}$, (\ref{2.13})$_{2}$ and (\ref{2.13})$_{3}$
must be proportional to each other. Given that the case, we
automatically have $m_{1}=m_{2}=m_{3}=m$, which corresponds to the
case of the three identical fullerenes we study in this paper. In
addition, we require $\beta_{1}=\beta_{2}=\beta_{3}=1/3$ resulting
to the three identical fullerenes allocated at the vertices of an
equilateral triangle, i.e., $r_{kj}=r$. We have

\begin{equation}
\frac{1}{m\omega^{2}}\frac{F(r)}{r}=\frac{1}{3}, \quad\quad\quad
\omega=\sqrt{\frac{3F(r)}{mr}}.\label{2.14}
\end{equation}
On the other hand, we can directly obtain the same result by
investigating the equation of motion given by Eq.~(\ref{2.7})$_{1}$.
Upon assuming $F(r_{12})= F(r_{12})=F(r_{13})=F(r)$ and
$m_{1}=m_{2}=m_{3}=m$, we deduce

\begin{equation}
\vec{\ddot{r}}=-\frac{2m}{m^{2}}\frac{F(r)}{r}\vec{r}+\frac{1}{m}\frac{F(r)}{r}\left(
r+\frac{m}{2m}-\frac{m}{2m}
\right)\vec{r}-\frac{1}{m}\frac{F(r)}{r}\left(
r+\frac{m}{2m}+\frac{m}{2m}
\right)\vec{r}=-\frac{3F(r)}{mr}\vec{r},\label{2.15}
\end{equation}
which represents a simple harmonic motion with angular velocity
$\omega$, given exactly by Eq.~(\ref{2.14})$_{2}$.

\begin{table}[h]
\begin{center}
\begin{tabular}{lll}
\hline\hline
Radius of C$_{60}$  & $a=3.55$ {\AA}  \\
Carbon-carbon bond length   & $\sigma = 1.421$ {\AA}  \\
Mean surface density of C$_{60}$   & $n_{f}= 0.3789$ {\AA}$^{-2}$ \\
Mass of a single carbon atom   & $m_{c}=1.993\times 10^{-26}$ kg  \\
Mass of a single C$_{60}$ fullerene  & $m=1.196 \times 10^{-24}$ kg \\
Attractive constant   &  $A = 17.4$ $\mbox{eV}\times$ {\AA}$^{6}$ \\
Repulsive constant   & $B = 29\times 10^{3}$ $\mbox{eV}\times${\AA}$^{12}$ \\
\hline\hline
\end{tabular}
\caption{\footnotesize{Numerical values of constants used in the
model}} \label{tb:Xname}
\end{center}
\end{table}

\begin{figure}[!]
  \centering
  \includegraphics[width=12cm,height=8cm]{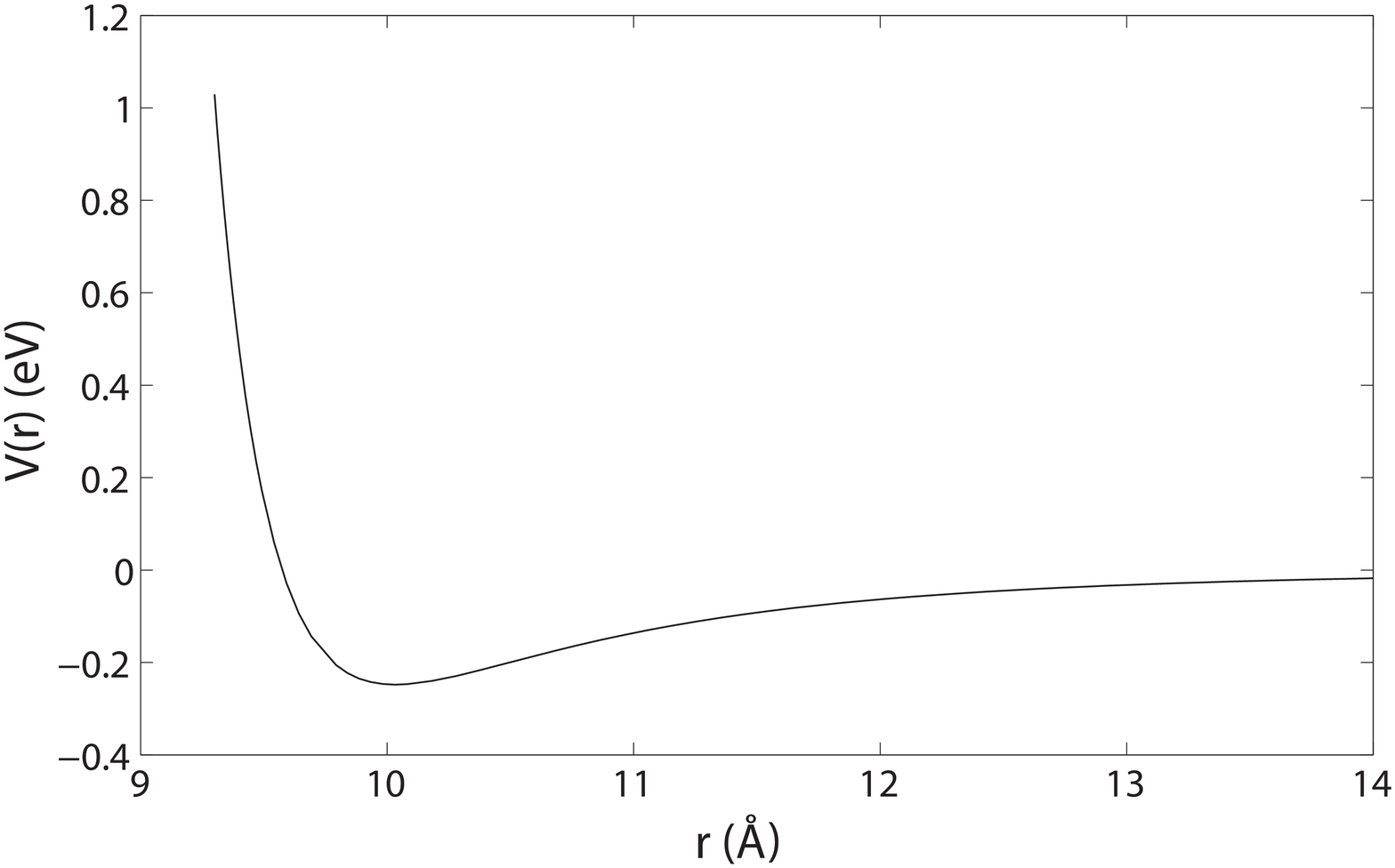}\\
  \caption{Total molecular energy $V(r)$ of two C$_{60}$ fullerenes}\label{fig:Fig3}
\end{figure}

Here, we carry out the theoretical results derived above. All the
necessary constants used in this paper can be found in
Table.~\ref{tb:Xname}. We note that while the molecular interactions
are attractive at long range due to the van der Waals forces, they
are repulsive at short range as the result of overlapping electron
orbitals.  The total pairwise molecular interactions between two
C$_{60}$ fullerenes, which is approximated by the continuous
approach, termed the total molecular energy, is shown in
Eq.~(\ref{2.2}) and is plotted in Fig.~\ref{fig:Fig3}. We find that
the most stable configuration for two C$_{60}$ fullerenes are at
$9.8$ {\AA} with a much higher energy well depth in comparison to
that of the two carbon atoms that is $0.28$ eV. Given the total
molecular energy, we can determine the total molecular forces
between two fullerenes by differentiating, and the force $F(r)$
given in Eq.~(\ref{2.4}). Here, we define positive forces to be
attractive forces and vice versa. Since the centripetal forces
between three fullerenes are provided by the attractive van der
Waals forces between molecules only, upon using Eq.~(\ref{2.14}), we
can predict the collective angular velocity $\omega$, which is
plotted in Fig.~\ref{fig:Fig5}. Since only attractive van der Waals
forces contribute to the orbiting behavior, no orbiting phenomenon
can be observed for $r<10$ {\AA}, where the van der Waals forces
become repulsive. Given the orbiting regimes, the maximum angular
velocity is found to occur at $0.12$ THz, which is substantially
higher than all gigahertz nanooscillators proposed in the existing
literature. As such the three C$_{60}$ fullerenes might be used as a
possible ultra high frequency nano-device.

\begin{figure}[!]
  \centering
  \includegraphics[width=12cm,height=8cm]{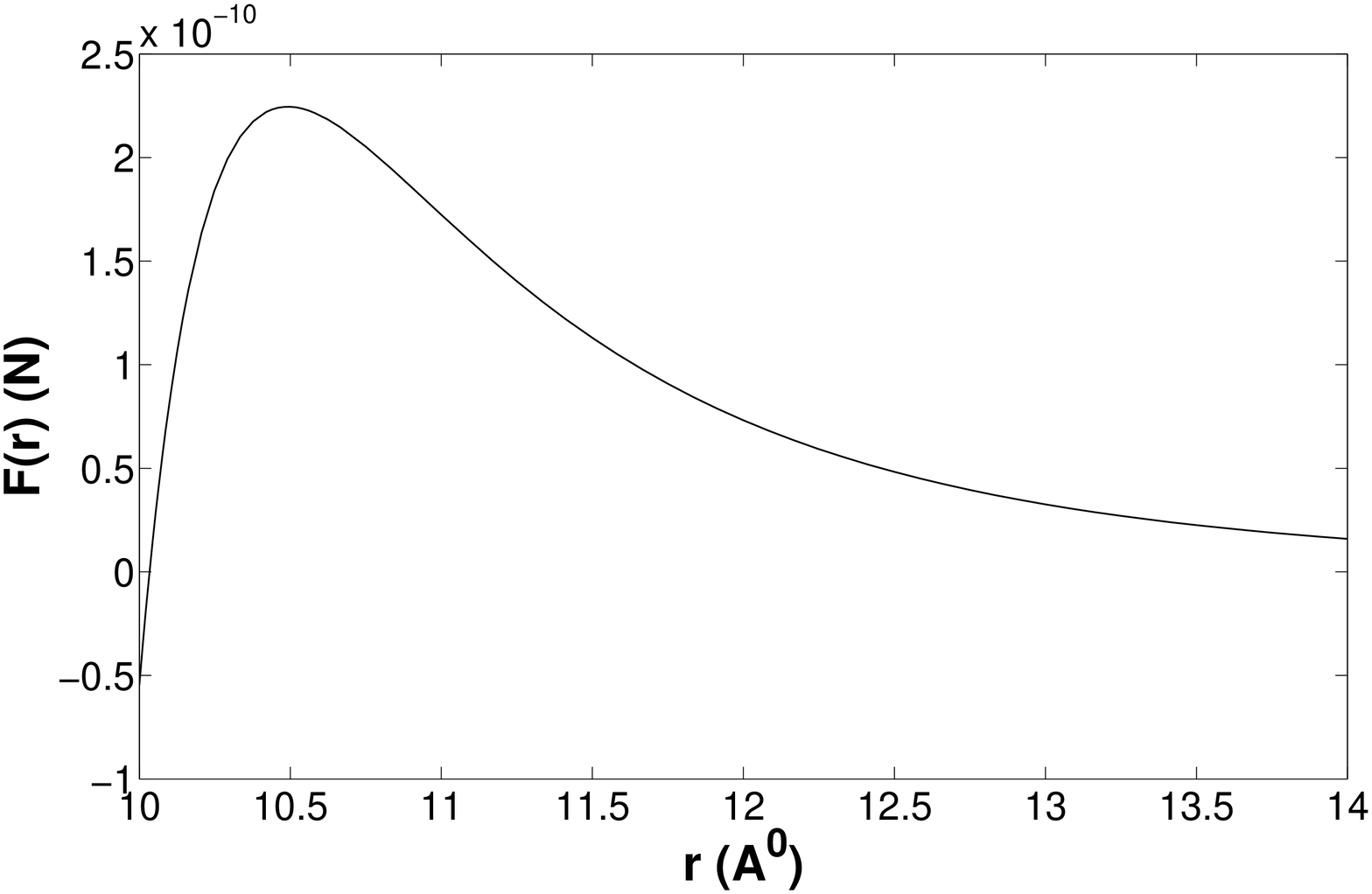}\\
  \caption{Total molecular force $F(r)$ acting on one fullerene}\label{fig:Fig4}
\end{figure}

\begin{figure}[!]
  \centering
  \includegraphics[width=12cm,height=8cm]{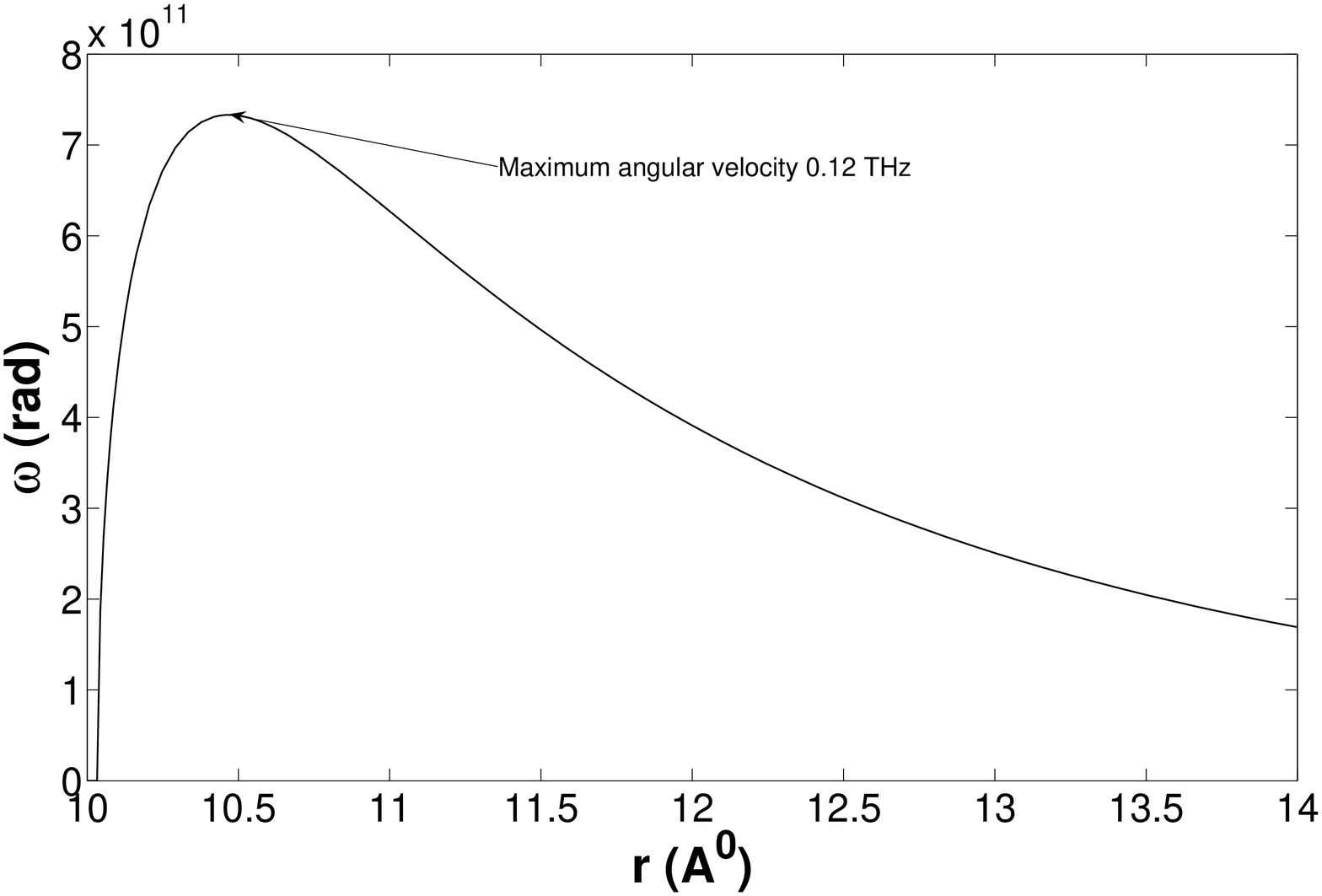}\\
  \caption{Collective angular velocity $\omega$ versus $r$, where the maximum angular velocity occurs at $0.12$ THz at approximately $r=10.5$ {\AA}}\label{fig:Fig5}
\end{figure}

\section{Capture of a carbon atom by two fullerenes}
\begin{figure}[!]
  \centering
  \includegraphics[width=12cm,height=8cm]{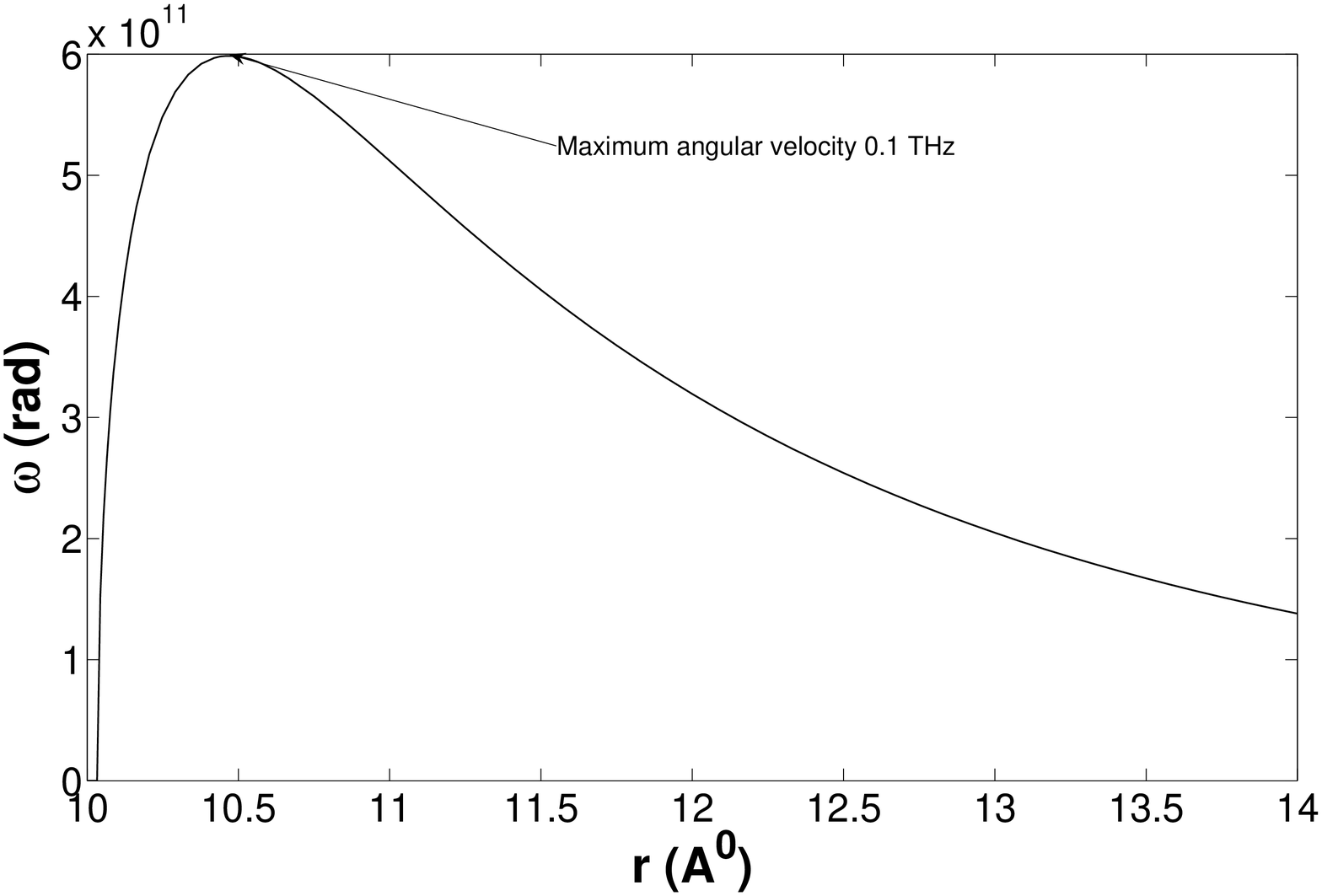}\\
  \caption{Collective angular velocity $\omega$ versus $r$, where the maximum angular velocity occurs at $0.1$ THz around $r=10.5$ {\AA}}\label{fig:Fig6}
\end{figure}

In this section, we investigate the motion of a carbon atom under
the influence of the potential energy from two C$_{60}$ fullerenes.
We assume that the molecular interactions of the carbon atom is so
small that it cannot influence the dynamics of the two C$_{60}$
fullerenes. From which, the Jacob coordinates, i.e. Eq.~(\ref{2.7})
reduce to

\begin{eqnarray}
&& \ddot{\vec{r}}=-\frac{2}{m}\frac{F(r_{12})}{r_{12}}\vec{r},
\nonumber\\
&&
\ddot{\vec{\rho}}=-\frac{1}{m_{c}}\left[\frac{F({\rho_{1}})}{\rho_{1}}(\vec{\rho}+\frac{1}{2}\vec{r})+\frac{F(\rho_{2})}{\rho_{2}}(\vec{\rho}-\frac{1}{2}\vec{r})\right],\label{3.1}
\end{eqnarray}
where $m$, $m_{c}$, $\rho_{1}$ and $\rho_{2}$ denote the mass of the
C$_{60}$ molecule, the mass of the carbon atom, the distance between
the atom and the first fullerene and the distance between the atom
and the second fullerene respectively. Eq.~(\ref{3.1})$_{1}$
constitutes the usual two body problem, which has the usual physical
interpretation. If we assume the same scenario as in Section 2 but
for the two fullerenes, that is the two fullerenes move collectively
in a circular orbit of radius $R$ in a plane and with the same
angular velocity $\omega$, Eq.~(\ref{3.1})$_{1}$ becomes simple
harmonic motion with angular velocity equal to $\sqrt{2F(r)/mr}$ for
a fixed $r$ and this is plotted in Fig.~\ref{fig:Fig6}. In the later
part of this section, we denote $R=10.5$ {\AA} to be the circular
radii, where the collective angular velocity of the two body systems
attaches maximum value.

Similarly, we assume that the center of mass of the two fullerenes
are located at the center of a cartesian coordinate system and we
introduce a rotating frame $(\xi,\eta)$ with the same angular rate
as the collective angular velocity $\omega$ so that the two
fullerenes remain stationary with respect to the rotating frame. The
position of the two fullerenes can then be determined by solving
$\xi_{1}+\xi_{2}=0$ and $\xi_{2}-\xi_{1}=R$ simultaneously and the
situation is shown schematically in Fig.\ref{fig:Fig6}. Since the
dynamics of the two body systems is determined, the only problem now
left is the motion of the carbon atom about the center of mass,
which has to be solved from Eq.~(\ref{3.1})$_{2}$.

\begin{figure}[!]
  \centering
  \includegraphics[width=12cm,height=10cm]{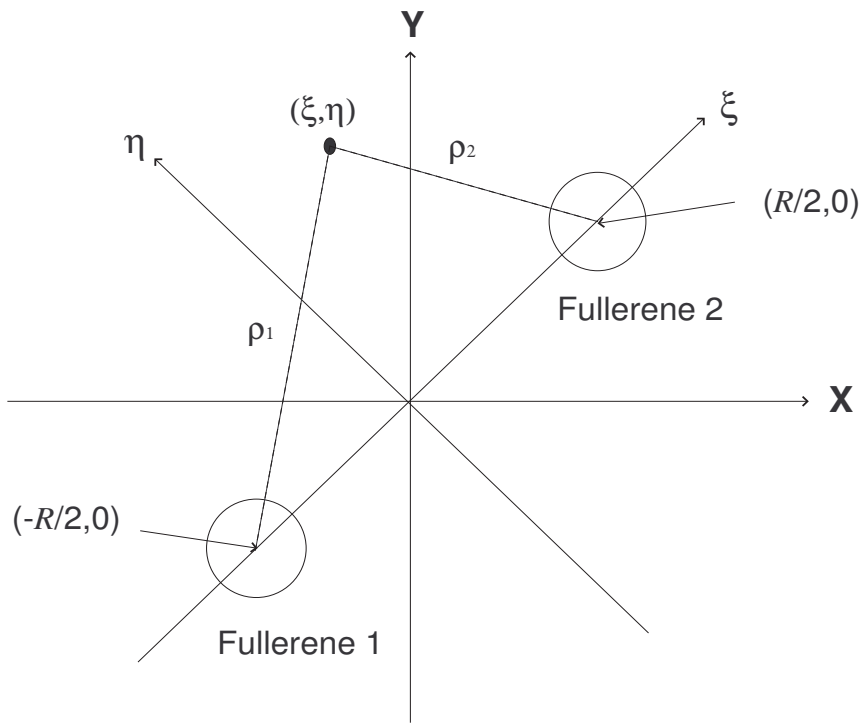}\\
  \caption{Capture of carbon atom by two identical C$_{60}$ fullerene}\label{fig:Fig7}
\end{figure}

Now making the same circular planar assumptions as above, the third
component, i.e the equations of motion of the carbon atom, i.e.
Eqs.~(\ref{2.10}) becomes

\begin{eqnarray}
&& \ddot{\xi}-2\omega \dot{\eta}-\omega^{2}\xi = \frac{1}{m_{c}}
\left\{
\frac{F(\rho_{1})}{\rho_{1}}(\xi_{1}-\xi_{3})+\frac{F(\rho_{2})}{\rho_{2}}(\xi_{2}-\xi_{3})
\right\}, \nonumber\\
&&\ddot{\eta}+2\omega\dot{\xi}-\omega^{2}\eta=\frac{1}{m_{c}}\left\{\frac{F(\rho_{1})}{\rho_{1}}(\eta_{1}-\eta_{3})+\frac{F(\rho_{2})}{\rho_{1}}(\eta_{2}-\eta_{3})\right\},\label{3.2}
\end{eqnarray}
where $m_{c}$, $\xi$, $\eta$, $\rho_{1}=\sqrt{(\xi
+R/2)^{2}+\eta^{2}}$ and $\rho_{2}=\sqrt{(\xi-R/2)^{2}+\eta^{2}}$
denote the mass of the carbon atom, $\xi_{3}$, $\eta_{3}$, $r_{13}$
and $r_{23}$ respectively. For simplicity, we only work on
Eq.~(\ref{3.2})$_{1}$ and Eq.~(\ref{3.2})$_{2}$ can be determined in
a very similar way. Starting from Eq.~(\ref{3.2})$_{1}$ we may
deduce

\begin{eqnarray}
\ddot{\xi}-2\omega\dot{\eta}-\omega^{2}\xi&=&\frac{1}{m_{c}}\left\{\frac{F(\rho_{1})}{\rho_{1}}
\left(-\frac{R}{2}-\xi \right)+\frac{F(\rho_{2})}{\rho_{2}}
\left(\frac{R}{2}-\xi \right) \right\} \nonumber\\
&=& -\frac{1}{m_{c}}\left\{\frac{F(\rho_{1})}{\rho_{1}}
\left(\xi+\frac{R}{2} \right)+\frac{F(\rho_{2})}{\rho_{2}} \left(\xi
-\frac{R}{2}\right) \right\} \nonumber\\
&=& -\frac{1}{m_{c}}\left\{ F(\rho_{1})\frac{\partial
\rho_{1}}{\partial \xi} + F(\rho_{2})\frac{\partial
\rho_{2}}{\partial \xi}\right\} \nonumber\\
&=& \frac{1}{m_{c}}\left\{ \frac{\partial V(\rho_{1})}{\partial
\rho_{1}}\frac{\partial \rho_{1}}{\partial \xi}+\frac{\partial
V(\rho_{2})}{\partial
\rho_{2}}\frac{\partial \rho_{2}}{\partial \xi} \right\} \nonumber\\
&=& \frac{1}{m_{c}}\left\{ \frac{\partial V(\rho)}{\partial \xi}
\right\}, \label{3.3}
\end{eqnarray}
where $V(\rho)=V(\rho_{1})+V(\rho_{2})$. We can further simplify
Eq.~(\ref{3.3}) to become

\begin{equation}
\ddot{\xi}-2\omega\dot{\eta}=\frac{1}{2m_{c}}\frac{\partial
\Phi}{\partial \xi}, \label{3.4}
\end{equation}
where $\Phi=\Phi_{1}+\Phi_{2}$, and
$\Phi_{1}=V(\rho)+m_{c}\omega^{2}\xi^{2}$ and
$\Phi_{2}=V(\rho)+m_{c}\omega^{2}\eta^{2}$. We therefore conclude
that

\begin{equation}
\ddot{\xi}-2\omega\dot{\eta}=\frac{1}{2m_{c}}\frac{\partial
\Phi}{\partial \xi}, \quad\quad\quad
\ddot{\eta}+2\omega\dot{\xi}=\frac{1}{2m_{c}}\frac{\partial
\Phi}{\partial \eta},\label{3.5}
\end{equation}
and Eq.~(\ref{3.5}) can then be recast in the form

\begin{equation}
\frac{d}{dt}\left[\frac{1}{2}m_{c}\left( \dot{\xi}^{2} +
\dot{\eta}^{2}\right) \right] = \frac{\partial \Phi}{\partial t}.
\label{3.6}
\end{equation}
and on integrating both sides, we obtain the energy integral

\begin{equation}
\frac{1}{2}m_{c}\left( \dot{\xi}^{2}+\dot{\eta}^{2} \right) = \Phi
+C\geq 0,\label{3.7}
\end{equation}
where $C$ denotes the Jacobi constant. As in the classical three
body problems, we are interested in determining where the carbon
atom is stationary with respect to the rotating frame under the
influence of the two fullerenes. In other words, we require the
velocity in both the $\xi$-direction and the $\eta$-direction to be
zero, i.e. $\dot{\xi}=\dot{\eta}=0$ and hence from Eq.~(\ref{3.5}),
we have $\partial \Phi/\partial \xi =
\partial \Phi/\partial \eta = 0$.

\begin{figure}[!]
\centering
  \includegraphics[width=12cm,height=8.5cm]{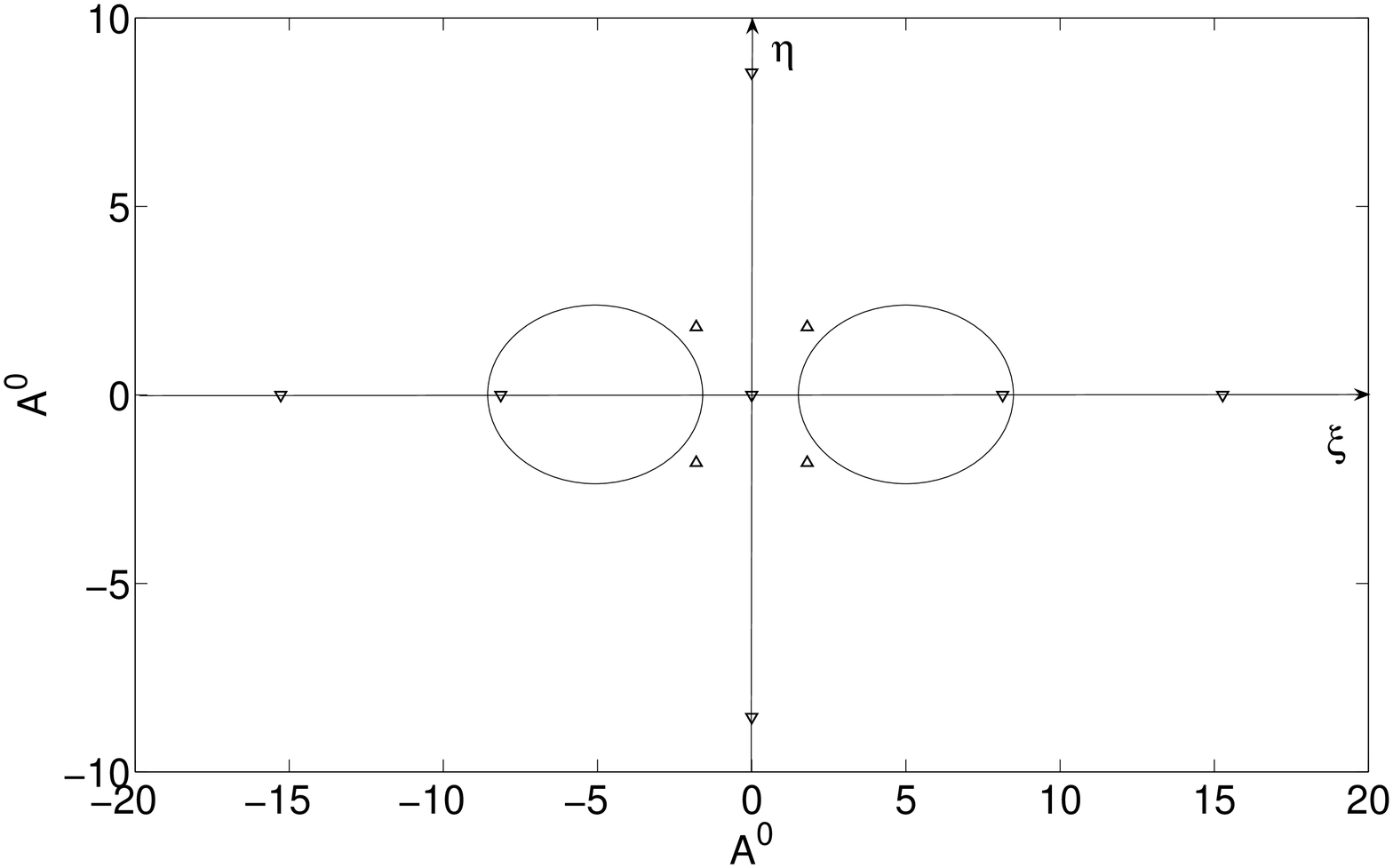}\\
    \caption{Stationary points of carbon atom with respect to the rotating frame $(\xi,\eta)$, which are marked by $\nabla$ and the maximum velocity of the carbon atom, which are marked by $\triangle$. The two circles represent the two C$_{60}$ fullerenes}\label{fig:Fig8}
\end{figure}

The positions, for which $\dot{\xi}=\dot{\eta}=0$ is satisfied, are
numerically obtained and shown in Fig.~\ref{fig:Fig8}. We note that
two minima are located inside the two fullerenes, which indicate the
encapsulation of the carbon atom inside the fullerenes and due to
the symmetry of the system, all minimum occur at either $\xi=0$ or
$\eta=0$. Since $C$ is an integral constant throughout the
conservative system, we can obtain the value of $C$ from a
particular case, the simplest one is $(\dot{\xi},\dot{\eta})=(0,0)$
when $(\xi,\eta)=(0,0)$. $C=-\Phi(\xi=0,\eta=0)$ is found to be
9$\times 10^{-18}$ J. Upon substituting $C$ back into
Eq.~(\ref{3.7}), we can determine the velocity profile of the carbon
atom and hence the maximum velocity of the carbon atom, which is
also shown in Fig.~\ref{fig:Fig8}.

\section{Conclusion}
We adopt the continuous approximation together with the
Lennard-Jones potential to model the van der Waals between
fullerenes. The total pairwise molecular interactions of the atoms
between two fullerenes, using molecular dynamics simulations usually
takes a substantial computational effort to implement. Here, such a
total molecular energy is computed analytically and the
corresponding total molecular force can be determined by
differentiating. For the restricted three body problems derived in
section 2, we can estimate the collective angular velocity of the
two and three fullerenes systems. In particular, the maximum angular
velocity reaches the terahertz range. In addition, we determine the
stationary points and the points with maximum velocity of the carbon
atom captured by two oscillating fullerenes.

\section*{acknowledgments}
We gratefully acknowledge the support from the Discovery Project
Scheme of the Australian Research Council and some very helpful
comments from Dr. Barry Cox.

\bibliography{3body}
\bibliographystyle{plain}

\end{document}